\documentclass[aps,prl,twocolumn,amsmath,showpacs,superscriptaddress,nofootinbib,preprintnumbers]{revtex4-1}

\usepackage{verbatim}
\usepackage[T1]{fontenc}
\usepackage[utf8]{inputenc}
\usepackage[american]{babel}
\usepackage{epsfig}
\usepackage{graphicx}
\usepackage{hyperref}

\usepackage{booktabs}
\usepackage{multirow}
\usepackage{dcolumn}
\usepackage{amsmath}
\usepackage{mathtools}
\usepackage{amsfonts}
\usepackage{amssymb}
\usepackage{epstopdf}
\usepackage{bm}
\usepackage{braket}
\usepackage{enumitem}
\usepackage{soul}
\usepackage[table]{xcolor}
\usepackage{color}
\usepackage{transparent}
\usepackage{pifont}

\definecolor{navyblue}{rgb}{0.0, 0.0, 0.5}
\definecolor{royalblue}{rgb}{0.25, 0.41, 0.88}
\definecolor{cadmiumgreen}{rgb}{0.0, 0.42, 0.24}
\definecolor{blue-violet}{rgb}{0.54, 0.17, 0.89}
\definecolor{darkviolet}{rgb}{0.58, 0.0, 0.83}
\definecolor{orange(colorwheel)}{rgb}{1.0, 0.5, 0.0}
\def\VEV#1{\left\langle #1 \right\rangle}
\usepackage{hyperref}
\hypersetup{
    colorlinks=true, % false: boxed links; true: colored links
    linkcolor=royalblue, % color of internal links (change box color with linkbordercolor)
    citecolor=magenta}

\newcommand\ee{\end{equation}}
\newcommand\be{\begin{equation}}
\newcommand\eea{\end{eqnarray}}
\newcommand\bea{\begin{eqnarray}}

\def\hatn{\mathbf{\hat n}}
\renewcommand{\vec}{\bm}

\usepackage{booktabs}
\usepackage{multirow}
\usepackage{dcolumn}
\usepackage{colortbl}
\newcommand\vertsp{\rule[-2mm]{1mm}{0mm} &}
\newcommand\horsp{\rule[-1.5mm]{0mm}{4.125mm}}
\newcommand\morehorsp{\rule[-2.25mm]{0mm}{6mm}}

\definecolor{magenta(process)}{rgb}{1.0, 0.0, 0.56}

\definecolor{darkspringgreen}{rgb}{0.09, 0.45, 0.27}

\definecolor{royalblue(web)}{rgb}{0.25, 0.41, 0.88}
\begin{document}

\title{Cosmic Birefringence Test of the Hubble Tension}

\author{Ludovico M. Capparelli}
\email{ludovico.mark.capparelli@roma1.infn.it}
\affiliation{Physics Department and INFN, Universit\`a di Roma ``La Sapienza'', Ple Aldo Moro 2, 00185, Rome, Italy}

\author{Robert R. Caldwell}
\email{robert.r.caldwell@dartmouth.edu}
\affiliation{Department of Physics \& Astronomy, Dartmouth College, Hanover, NH 03755 USA}

\author{Alessandro Melchiorri}
\email{alessandro.melchiorri@roma1.infn.it}
\affiliation{Physics Department and INFN, Universit\`a di Roma ``La Sapienza'', Ple Aldo Moro 2, 00185, Rome, Italy}

\date{\today}

%\preprint{}

%%%%%%%%%%%%%%%%%%%%%%%%%%%%%%%%%%%%%%%%%%%%%%%%%%%%%%%%%%%%%%%%%%%%%
\begin{abstract}
An early dark energy component consisting of a cosmic pseudo Nambu-Goldstone boson has been recently proposed to resolve the Hubble tension -- the four-sigma discrepancy between precision measurements of the expansion rate of the universe. Here we point out that such an axion-like component may be expected to couple to electromagnetism by a Chern-Simons term, and will thereby induce an anisotropic cosmic birefringence signal in the polarization of the cosmic microwave background (CMB). We show that observations of the rotation-angle power spectrum and cross-correlation with CMB temperature anisotropy can confirm the presence of this early dark energy component. Future CMB data as expected from the CMB-S4 experiment will improve sensitivity to this effect by two orders of magnitude and help in discriminating between different Hubble tension scenarios.
\end{abstract}

\maketitle

%%%%%%%%%%%%%%%%%%%%%%%%%%%%%%%%%%%%%%%%%%%%%%%%%%%%%%%%%%%%%%%%%%%%%
{\it Introduction.}---The Planck experiment has provided precise measurements of the Cosmic Microwave Background (CMB) anisotropies \cite{Aghanim:2018eyx}. While in excellent agreement with the predictions of the standard cosmological model, the new data present several interesting tensions and anomalies that could hint for the presence of new physics. The most statistically significant tension concerns the value of the Hubble constant. The Planck experiment infers a value of the Hubble constant of $H_0=67.4\pm0.5$ km/s/Mpc at $68 \%$ C.L., assuming the standard $\Lambda$CDM model. Yet this value is discordant at more than four standard deviations with the recent, direct $H_0$ determination $H_0=74.03\pm 1.42$ km/s/Mpc at $68 \%$ C.L. from Riess {\it et al.} (2019) \cite{Riess:2019cxk}, based on local measurements of the distance ladder. This tension has put the standard cosmological model under intense scrutiny \cite{Freedman:2017yms}.

A variety of physical mechanisms have been proposed to alleviate the tension \cite{Zhao:2017cud,DiValentino:2017zyq,Poulin:2018zxs,DiValentino:2017oaw,DiValentino:2017iww,Vattis:2019efj,Pandey:2019plg,Knox:2019rjx,Adhikari:2019fvb}.
One proposal \cite{Poulin:2018cxd,Agrawal:2019lmo} suggests that a species of {\it early dark energy} (EDE) is responsible for the shift in recombination-era length scales needed to raise the inferred value of the Hubble constant. These models are attractive since they also provide an excellent fit to Baryonic Acoustic Oscillations and Supernovae type Ia data. Future measurements of CMB anisotropies and large scale structure will refine the Hubble tension and significantly constrain these models \cite{Smith:2019ihp}.

Here we point out that a leading class of EDE models generically predicts a unique observable that can provide a critical test of these models. Specifically, EDE in the form of a cosmic pseudo Nambu-Goldstone boson (PNGB) may be expected to couple to electromagnetism (EM) by a Chern-Simons term. This interaction leads to Cosmic Birefringence (CB) in the polarization of the CMB.  A detection of CB by future CMB experiments could, therefore, provide a smoking gun for cosmic fields and insight into fundamental physics.

CB causes the anomalous rotation of the plane of polarization as CMB photons travel to us from the surface of last scattering \cite{Carroll:1989vb, Lue:1998mq}. This rotation converts the E-mode pattern into B-modes, and vice versa. Since the power spectrum of the E-mode pattern dominates over the B-mode, CB essentially produces a transfer of primordial CMB E-modes into B-modes. A measurement of the rotation angle $\alpha$ is actively sought, and there are already tight constraints. The current bound is $\langle\alpha^2\rangle^{1/2} \lesssim 0.5^\circ$ \cite{Molinari:2016xsy,Contreras:2017sgi}. Absolute calibration of the polarization detectors, however, is a major source of systematic uncertainty and a severe limiting factor in attempts to detect CB \cite{Pogosian:2019jbt}. 

Another possibility is to seek the anisotropies in the CB angle. The measurement of higher-order temperature and polarization correlations can be used to reconstruct the rotation angle $\alpha$ as a function of position on the sky \cite{Kamionkowski:2008fp,Gluscevic:2009mm}. This observable, while smaller in amplitude, is less affected by experimental systematics. Moreover, anisotropies of the rotation angle can provide key information about the physics underlying the CB. Indeed, if CB is due to an EM Chern-Simons-coupled scalar field, then fluctuations in the scalar field will result in CB anisotropies over the sky \cite{Pospelov:2008gg,Kamionkowski:2008fp,Gluscevic:2009mm,Yadav:2009eb}. The imprint of the scalar field will be present in the power spectrum of anisotropies of the rotation angle, and correlations with the temperature pattern \cite{Caldwell:2011pu}. That is our smoking gun.

It is timely, therefore, to investigate the CB rotation angle anisotropies that are predicted in the proposed Hubble-tension solutions. As we show, fluctuations of the scalar field leave a distinct imprint that may be within reach of ongoing and next-generation CMB experiments such as the Simons Observatory \cite{Ade:2018sbj}, CMB-S4 \cite{Abazajian:2016yjj}, and LiteBIRD \cite{Matsumura:2013aja,Suzuki:2018cuy}.

%%%%%%%%%%%%%%%%%%%%%%%%%%%%%%%%%%%%%%%%%%%%%%%%%%%%%%%%%%%%%%%%%%%%%
{\it EDE solution to the Hubble tension.}---A model of EDE from the string axiverse was recently proposed as a solution the Hubble tension \cite{Poulin:2018cxd}. In this scenario, EDE is a PNGB $\phi$ with potential
\begin{equation}
    V_n(\phi) = \Lambda^4 (1 - \cos \frac{\phi}{f})^{n}.
\end{equation}
Such a potential arises non-perturbatively in string axiverse scenarios \cite{Arvanitaki:2009fg,Kamionkowski:2014zda}. In the standard formulation, the PNGB is a spin-0 degree of freedom that emerges from the spontaneous breakdown of a continuous symmetry at scale $f$ which is also explicitly broken by the term $\Lambda$. This construction is ideal for dark sector physics, e.g. dark energy, EDE, and dark matter; long range interactions are suppressed by a shift symmetry $\phi\to \phi$+constant, and the small value of $\Lambda$ desired for dark energy is technically natural \cite{Frieman:1995pm,Carroll:1998zi}. For $n=1$, we obtain the usual form for the axion potential, whereas $n>1$ can arise from higher-order instanton corrections \cite{Arvanitaki:2009fg,Kamionkowski:2014zda}. It is useful to define a mass $m\equiv{\Lambda^2}/{f}$, although we take care to note that the leading term at the potential minimum is $\propto \phi^{2n}$. Here, we fix $f=M_{Pl}$ where $M_{Pl}$ is the Planck mass. 

To model the evolution, we split the field into a homogeneous component plus a small perturbation: $\phi(\vec{x}, \tau) = \phi_0(\tau) + \delta \phi (\vec{x}, \tau)$. The background equation is
\begin{equation}
    \ddot{\phi}_0 + 2 \mathcal{H} \dot{\phi}_0 + a^2 V'_0 = 0,
\end{equation}
where the dots denote derivatives with respect to conformal time, $\mathcal{H}=\dot{a}/a$ is the conformal Hubble parameter, and $'$ denotes derivatives with respect to $\phi$. We give the equation of motion for the Fourier transform of the perturbation with wavenumber $k$ in two coordinate gauges, first in conformal-Newtonian gauge with potentials $\Phi,\,\Psi$
\begin{equation}
  \delta \ddot{\phi} + 2\mathcal{H} \delta\dot{\phi} + a^2 V''_0 \delta \phi + k^2 \delta \phi =  \dot{\phi}_0(3\dot\Phi+\dot\Psi)-2 a^2 V'_0 \Psi,
  \label{eqn:perturbationphinewton}
\end{equation}
and in the synchronous gauge with potential $h$
\begin{equation}
  \delta \ddot{\phi} + 2\mathcal{H} \delta\dot{\phi} + a^2 V''_0 \delta \phi + k^2 \delta \phi =  -\frac{1}{2}\dot{h}\dot{\phi}_0.
  \label{eqn:perturbationphisync}
\end{equation}
We use the same notation and conventions for the gauges as in Ref.~\cite{Ma:1995ey}.
 
The evolution of the homogeneous field passes through different stages. Initially, the field is frozen at its starting value of $\phi_{0i}$ due to Hubble friction. When the expansion slows enough, the field ``thaws'' and begins to roll downhill. Following \cite{Poulin:2018cxd}, we introduce $f_{EDE}(a) = \Omega_\phi (a) / \Omega_{tot}(a)$ to track the EDE abundance, and we define $a_c$ to be the scale factor where $\phi_0 = \frac{7}{8} \phi_{0i}$, the approximate point at which the field becomes dynamical. Before $a_c$, the field acts as a dark energy component to the universe with $w=-1$. After $a_c$ the field rolls downslope.

In the case of a power-law potential $V \propto \phi^{2 n}$ with $n \le (3+w_B)/(1-w_B)$, where $w_B$ is the background equation of state, the field eventually begins to oscillate around the minimum when the Hubble rate drops below the effective mass. For fast oscillations, the averaged equation of state is $\langle w \rangle = (n-1)/(n+1)$ and the effect of the field is similar to that of a fluid with the same equation of state. Hence, for $n=1$ the field eventually decays as matter, and for $n=2$ the field decays as radiation. The case $n > (3+w_B)/(1-w_B)$ is unusual: there is a stable attractor solution whereby the field rolls $\ln\phi_0 \propto t$ without reaching the bottom, and the equation of state evolves toward $w \to (1+n w_B)/(n-1)$ \cite{Liddle:1998xm}. (These two solutions describe the ``rock'' and ``roll'' of Ref.~\cite{Agrawal:2019lmo}.) The results similarly apply to the PNGB, for which the leading term at the potential minimum is a power law. 

The effects on the CMB are studied by solving the usual Boltzmann equations coupled with the equations for scalar field perturbation. Assuming that the field $\phi_0$ is frozen deep in the radiation era, then adiabatic initial conditions simply require that the perturbations and their derivatives vanish in the synchronous gauge. The initial conditions in the conformal-Newtonian gauge are found by changing gauges: $\delta \phi_c = \delta \phi_s + \alpha_s \dot{\phi}(\tau)$ where $\alpha_s = (\dot{h}+6\dot{\eta})/(2k^2)$ \cite{Ma:1995ey}. The coupled Boltzmann equations are solved using the Cosmic Linear Anisotropy Solving System (CLASS) numerical code~\cite{Blas:2011rf}. One challenge in integrating a scalar field is that the oscillations may have a period much shorter than a Hubble time. We have modified the CLASS code to ensure that the time-step is always much shorter than the period to accurately resolve the field evolution. 

An early dark phase with $n \ge 2$ and suitable mass parameters has been shown to alleviate the Hubble tension \cite{Poulin:2018cxd,Agrawal:2019lmo}. We have first reproduced these results by computing the theoretical CMB angular spectra for scalar fields with potentials with $n=2$ and $n=3$. We find that for $n=3$ and $m\simeq 3\cdot 10^{-26}$~eV, we obtain a good fit to the CMB Planck temperature and polarization anisotropy spectra. The best-fit value of the Hubble constant shifts to $H_0 = 71.9$~km/s/Mpc, which is now in better agreement with the Riess {\it et al.} (2019) value \cite{Riess:2019cxk}. In this model we obtain $log_{10}(a_c) = -3.7$ and $f_{EDE}(a_c)=0.06$, in agreement with Table 2 of Ref.~\cite{Poulin:2018cxd}. We have also investigated the case $n=2$, using $m\simeq 8\cdot 10^{-27}$~eV; the Hubble constant is $H_0 = 70.0$~km/s/Mpc and we obtain $log_{10}(a_c) = -3.7$ and $f_{EDE}(a_c) = 0.04$, also in agreement with Ref.~\cite{Poulin:2018cxd}.

%%%%%%%%%%%%%%%%%%%%%%%%%%%%%%%%%%%%%%%%%%%%%%%%%%%%%%%%%%%%%%%%%%%%%
{\it CB anisotropy from EDE.}---PNGBs remain dark thanks to the shift symmetry that prohibits most couplings to the Standard Model \cite{Carroll:1998zi}. A coupling to the Chern-Simons scalar of a gauge field respects this symmetry, however, and must be considered. The most significant such coupling is to EM, whereby the Lagrangian includes
\begin{equation}
\label{eq:FFdual}
     L \ni
      \frac{1}{4} G_{\gamma\phi} \phi F^{\mu\nu} \widetilde{F}_{\mu\nu},
\end{equation}
$\widetilde{F}_{\mu\nu} = \frac{1}{2}\epsilon_{\mu\nu\rho\sigma} F^{\rho\sigma}$ is the dual Faraday tensor, and $\epsilon_{\mu\nu\rho\sigma}$ is the Levi-Civita tensor. The dimensionful coupling $G_{\gamma\phi}$ depends on the specific model and is treated as as an unknown parameter to be measured, although we may reasonably expect it to be inversely related to the symmetry breaking scale. This interaction is responsible for CB \cite{Carroll:1989vb, Lue:1998mq}.
  
Linearly polarized light propagating over cosmological distances in a background field $\phi_0$ will undergo a global rotation of the polarization pseudovector. The rotation angle $\alpha$ is equal to
\begin{eqnarray}
     \alpha &=& \frac{1}{2} G_{\gamma\phi}\int d\tau
     \left(\frac{\partial}{\partial\tau} -
     \hatn\cdot\vec\nabla\right)\phi_0(\tau)  \cr
    &=& \frac{1}{2} G_{\gamma\phi} \Delta\phi_0,
\label{eqn:rotationangle}
\end{eqnarray}
where the integral is along the null path, $\tau$ is conformal time, and  $\Delta\phi_0$ is the change in the PNGB over the photon path. However, photons of the CMB are not all emitted at the same time. Rather, they are statistically distributed over the photon visibility function $g(\tau) = -\dot{\kappa} e^{-\kappa}$, with $\kappa$ being 
\begin{equation}
    \kappa = \int^{\tau_0}_{\tau} d\tau' a \sigma_T n_e,
\end{equation}
with $\tau_0$ being the conformal time today, $\sigma_T$ the Thompson cross section and $n_e$ the electron density. The uniform CB rotation angle can be expressed as
\begin{equation}
    \alpha = \frac{1}{2}G_{\gamma\phi}\phi_{obs} + \bar\alpha, \quad
    \bar\alpha = -\frac{1}{2}G_{\gamma\phi}\int^{\tau_0}_0 d\tau g(\tau) \phi_0(\tau), 
\end{equation}
with $\phi_{obs}$ the amplitude of the field at the observer. In the ``sudden'' approximation, in which the last scattering is instantaneous, $g(\tau)$ becomes a delta function and the idealization represented by Eq.~(\ref{eqn:rotationangle}) is recovered.

Anisotropies in the rotation angle on the sky arise from both the fluctuations of the scalar field and the visibility function. One way to appreciate the relevance of this effect is to consider a coordinate gauge in which the scalar field fluctuations vanish; anisotropies of the rotation angle are physical and cannot disappear, so there must be a compensating source of fluctuations. To that end, it is necessary to perturb $\bar\alpha$. As a first step we can unpack the integral to show that it is the solution to a Boltzmann-like differential equation along the line of sight,
\begin{equation}
\label{eqn:rotationEquationOrderZero}
    \dot{\bar{\alpha}} + a n_e \sigma_T \bar{\alpha} = -\frac{1}{2}G_{\gamma\phi} a n_e \sigma_T \phi_0,
\end{equation}
which can be derived from a phenomenological, covariant conservation equation $\nabla_\mu \bar{\alpha} - \sigma_T j_\mu \bar{\alpha} = \frac{1}{2}G_{\gamma\phi}\sigma_T j_\mu \phi_0$. Working with a spacetime metric $g_{\mu\nu} = a^2(\tau){\rm diag}(-1,1,1,1)$, in the source rest frame such that the current is $j^\mu=\frac{1}{a}(n_e,0,0,0)$, then Eq.~(\ref{eqn:rotationEquationOrderZero}) is recovered. Now it is straightforward to choose a gauge and perturb the covariant equation: $\phi_0\rightarrow \phi_0+\delta\phi$, $\bar{\alpha}\rightarrow \bar{\alpha} + \delta \alpha$ and $n_e\rightarrow n_e + \delta n_e$. The resulting linearized equation of motion in the synchronous gauge is
\begin{multline}
       \delta \dot{\alpha} + a \sigma_T n_e \delta \alpha + a \sigma_T \delta n_e \bar{\alpha} = \\ -\frac{1}{2} G_{\gamma\phi} a \sigma_T (n_e \delta \phi + \delta n_e \phi_0),
    \label{eqn:rotationSynchronous}
\end{multline}
integrated along the photon path. This is precisely what one would naively guess by perturbing Eq.~(\ref{eqn:rotationEquationOrderZero}). A similar equation is derived in the conformal-Newtonian gauge,
\begin{multline}
       \delta \dot{\alpha} + a \sigma_T n_e \delta \alpha =\\ -\frac{1}{2}G_{\gamma\phi} a \sigma_T n_e \delta \phi - a\sigma_T (\bar{\alpha} + \frac{1}{2} G_{\gamma\phi}\phi_0) (2\Psi n_e + \delta n_e).
\end{multline}
The equation for $\delta \alpha$ may be recast in integral form  
\begin{equation}
    \delta \alpha (\tau_0, \mathbf{k}) =\int^{\tau_0}_0 S_\alpha (\tau,\mathbf{k}) d\tau,
\end{equation}
where the source in the conformal-Newtonian gauge is given by 
\begin{equation}
    S_\alpha (\mathbf{k},\tau) = -g(\tau) \left ( \frac{1}{2}G_{\gamma\phi} \delta \phi + (\bar\alpha + \frac{1}{2}G_{\gamma\phi} \phi_0 )(2\Psi + \delta_b) \right ).
\label{eqn:alphaSource}
\end{equation}
We note that it is necessary to evolve equations for both $\bar\alpha$ and $\delta\alpha$ together. Also, fluctuations of the electron density are not tracked in a typical CMB Boltzmann code. We suggest the approximation $\delta n_e \simeq \delta_b n_e$, as we have done in the last equation above, which should be valid at the relevant times when baryons are non-relativistic. The above expressions are among our main results, which now take into account fluctuations in the distance to last scattering, thereby affecting the amount of polarization rotation that takes place along different lines of sight. 

To extract the observable rotation spectra we introduce the quantity
\begin{equation}
    \Delta_{\alpha,L}(k) = \int d\tau j_L(k|\tau_0-\tau|) S_\alpha (k,\tau)
\end{equation}
so that the rotation-angle power spectrum is given by
\begin{equation}
    C^{\alpha \alpha}_L = 4\pi \int \frac{k^2 dk}{2\pi^2} P_\Psi(k) \Delta_{\alpha,L}^2,
\end{equation}
where $P_\Psi(k)$ is the power spectrum of primordial adiabatic fluctuations of the gravitational potential. A cross-correlation between the rotation angle and the temperature also arises, 
\begin{equation}
    C^{\alpha T}_L = 4\pi \int \frac{k^2 dk}{2\pi^2} P_\Psi(k) \Delta_{T,L}(k) \Delta_{\alpha,L} (k),
\end{equation}
since the perturbation $\delta \phi$ is sourced by the same gravitational field responsible for the primary anisotropy (see e.g. \cite{Caldwell:2011pu}).

\begin{figure}[!htbp]
\centering
\includegraphics[width=.46\textwidth]{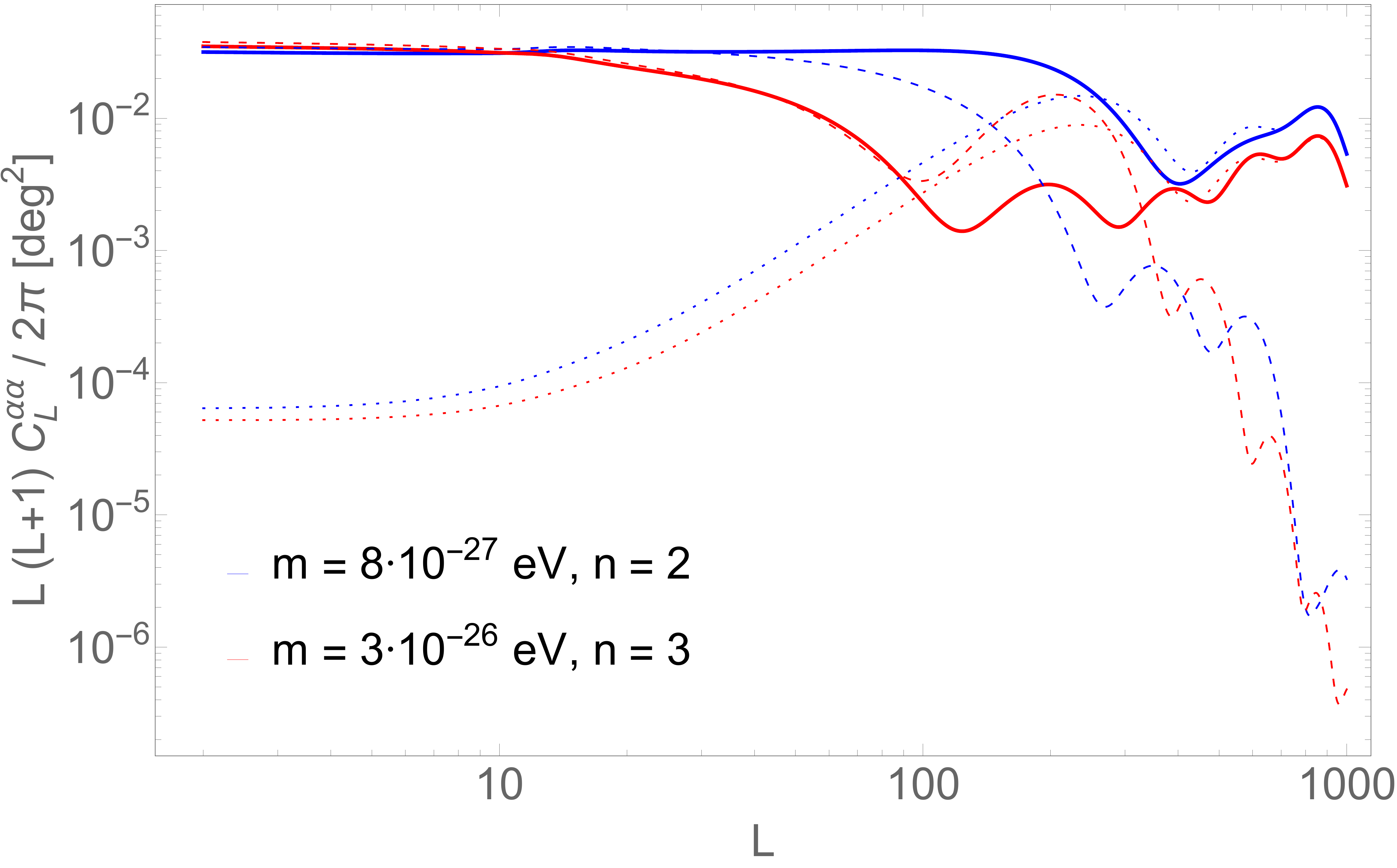}
\hbox{\hspace{+0.8em}\includegraphics[width=.45\textwidth]{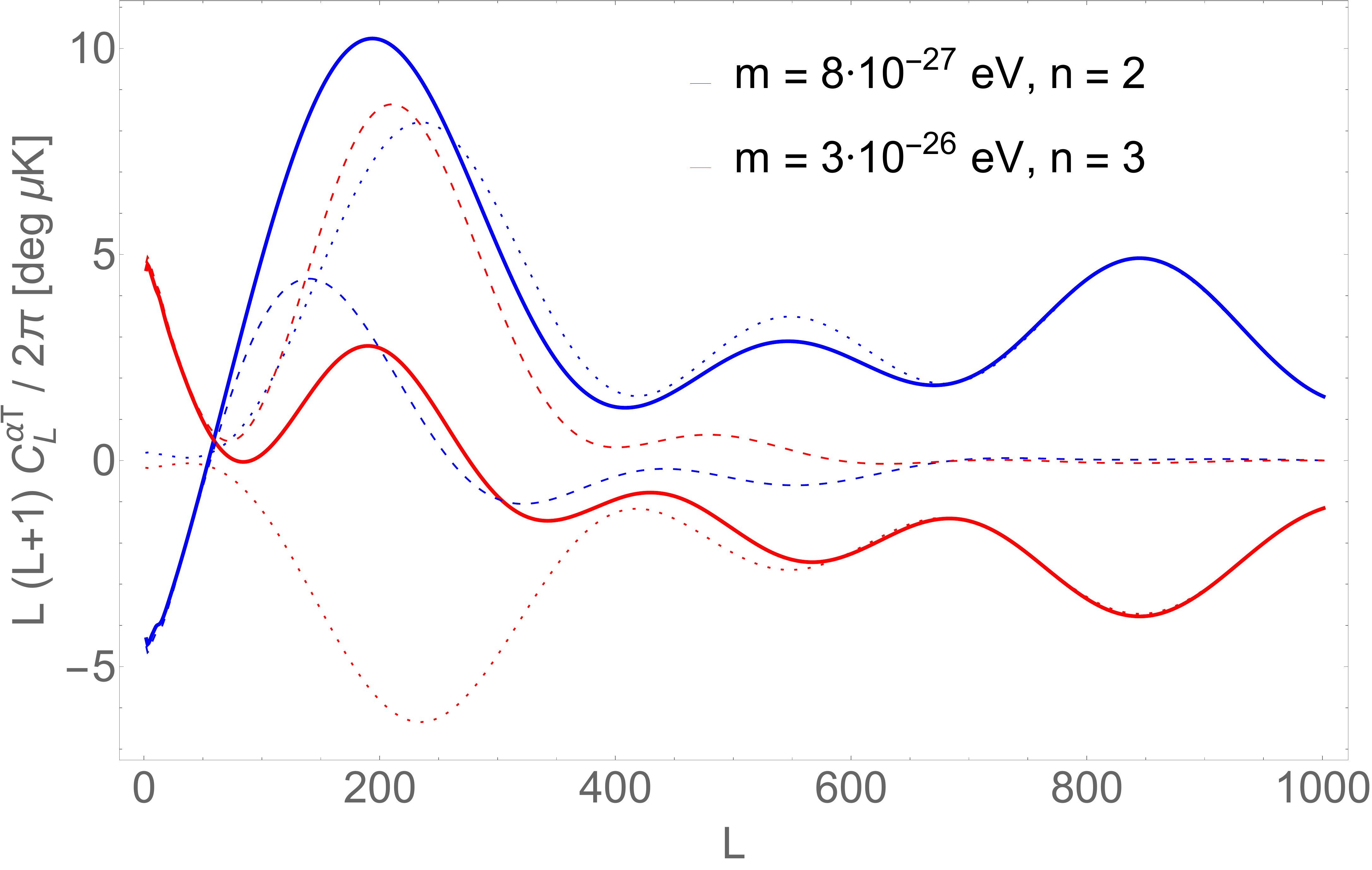}}
\caption{The rotation angle $\alpha \alpha$ (top) and $\alpha T$ (bottom) cross correlation spectra for two models of EDE that solve the Hubble tension problem are shown. Both cases have $G_{\gamma\phi} = 10^{-15}~{\rm GeV}^{-1}$. The solid curves are the total observable spectra. The dashed curves correspond to taking \emph{only} scalar field fluctuations contributions and the dotted curves are only density fluctuation contributions }
    \label{fig:spectra}
\end{figure}

We now examine the anisotropic CB arising in two different models proposed to solve the Hubble tension \cite{Poulin:2018cxd}. We calculated the $\alpha\alpha$ and $\alpha T$ spectra using a modified version of the Boltzmann code CLASS. The strikingly different results are shown in Fig.~\ref{fig:spectra}.

Both the EDE models predict $TT$, $TE$ and $EE$ spectra that are similar to each other and the $\Lambda$CDM model, within experimental error. However, the $\alpha\alpha$ and $\alpha T$ spectra are significantly different in shape, for the same amplitude of $G_{\gamma\phi}$. Thus the rotation spectra may be a powerful tool not only to confirm the presence of a cosmic scalar field but to discriminate between models. 

To understand the relevant contributions to the spectra, we have repeated the calculation assuming only contributions by fluctuations of the scalar field, or only baryon density and gravitational potential perturbations. We find that for $L \lesssim 50$ the scalar field fluctuations dominate the spectra, while for larger $L$ the density fluctuations become increasingly dominant. Indeed at high $L$ the peaks and troughs in the spectra are due to the well-known baryon acoustic oscillations. The amplitude at large $L$ is controlled by the value of $G_{\gamma \phi} \phi_0$ around recombination. We note that the overall sign of the $\alpha T$ spectra is arbitrary, depending on the signs of the coupling and the initial value of the background field $\phi_0$.

%%%%%%%%%%%%%%%%%%%%%%%%%%%%%%%%%%%%%%%%%%%%%%%%%%%%%%%%%%%%%%%%%%%%%
{\it Forecast for future experiments.}---Here we examine the ability of future CMB experiments to detect a cosmological birefringence signal given the proposed models by \cite{Poulin:2018cxd,Smith:2019ihp} and a coupling $G_{\gamma\phi}$. We consider four idealized experimental configurations, in that we neglect foregrounds and weak lensing, corresponding to the LiteBIRD satellite experiment (see \cite{Matsumura:2013aja,Suzuki:2018cuy}), the Stage-III experiment in both ``deep'' and ``wide'' surveys modes \cite{Allison:2015qca}, the Simons Observatory in the Small Aperture Telescopes (SAT) and Large Aperture Telescope (LAT) configurations \cite{Ade:2018sbj}, and the CMB-S4 ground-based telescope (see \cite{Abazajian:2016yjj}). The corresponding experimental specifications are listed in Table~\ref{tab:spec}.  

\begin{table}
\begin{center}
\begin{tabular}{lccc}
\toprule
\horsp
Experiment \vertsp Beam $\theta$ \vertsp Power noise $w^{-1/2}$\vertsp $f_{sky}$\\
&  &[\footnotesize$\mu$K-arcmin] & \\
\hline
\morehorsp
LiteBIRD      & $30$' & $4.5$& $0.7$\\
\morehorsp
S3deep      & $1$' & $4$ &$0.06$\\
\morehorsp
S3wide      & $1.4$' & $8$ &$0.4$\\
\morehorsp
Simons Obs. SAT      & $17$' & $2$ &$0.1$\\
\morehorsp
Simons Obs. LAT      & $1.4$' & $6$ &$0.4$\\
\morehorsp
CMB-S4      & $3$' & $1$&$0.4$\\
\bottomrule
\end{tabular}
\end{center}
\caption{Specifications for the different experimental configurations considered in our paper. In case of polarization spectra the noise $w^{-1}$ is multiplied by a factor 2.}
\label{tab:spec}
\end{table}

Following \cite{Caldwell:2011pu} we can derive the following expression for the $C_L^{\alpha\alpha}$ noise
power spectrum:
\begin{eqnarray}
    C_L^{\alpha\alpha,\mathrm{noise}} &\equiv& \VEV{ |\widehat
    \alpha_{LM}|^2}  \cr
    &=& \left[ \sum_{ll'} 
     \frac{(2l+1)(2l'+1)(F^{L,BE}_{ll'})^2}{ 4 \pi C^{BB\text{,
     map}}_l C^{EE \text{, map}}_{l'}} \right]^{-1}.
\label{eqn:alpha_variance}
\end{eqnarray}
where 
\begin{equation}
     C^{XX'\text{, map}}_l\equiv C^{XX'}_l|W_l|^2 +
     C^{XX'\text{, noise}}_l,
\label{Cobs}
\end{equation}
where $XX'\in \{TT, EE, BB, ET, EB, TB\}$ and $W_l(\theta) \equiv \exp\left[-l^2 \theta^2/(16\ln 2)\right]$ is the experimental window function where $\theta$ is the experimental beam as reported in Table~\ref{tab:spec}. The noise power spectra are
\begin{equation}
\begin{gathered}
  C^{TT\text{, noise}}_l \equiv f_\text{sky}w^{-1},\\
  C^{EE\text{, noise}}_l=C^{BB\text{, noise}}_l\equiv 2C^{TT\text{, noise}}_l,\\ 
  C^{EB\text{, noise}}_l=C^{TB\text{, noise}}_l\equiv 0,
\end{gathered} 
\label{Cnoise}
\end{equation}
where $w^{-1/2}$ is the power noise and $f_\text{sky}$ is the experimental sampled sky fraction. (See Table ~\ref{tab:spec}.) Finally, $F^{L,BE}_{ll'}$ is given by (see \cite{Caldwell:2011pu}):
\begin{equation}
	F^{L, BE}_{ll'}\equiv 2C^{EE}_{l'} 
\left( {\begin{array}{*{20}c}
           l & L & l'  \\
              2 & 0 & -2  \\
              \end{array}} \right) W_lW_{l'}, \quad
	F^{L, EB}_{ll'}\equiv F^{L, BE}_{l'l} ,
\end{equation}
where the objects in parentheses are Wigner-3j symbols. The expression for the variance of $\widehat C^{\alpha \alpha}_L$ is given by \cite{Caldwell:2011pu}:
\begin{equation}
     \left( \Delta \widehat C_L^{\alpha\alpha} \right)^2 \simeq
     \frac{2}{f_\text{sky}(2L+1)} \left(C_L^{\alpha\alpha,\mathrm{noise}}
     \right)^2,
\label{eqn:Calphaalpha_variance}
\end{equation}
while for the variance of $\widehat C^{\alpha T}_L$   we have:
\begin{equation}
     \left( \Delta C_L^{\alpha T} \right)^2 \simeq
     \frac{1}{f_\text{sky}(2L+1)} C_L^{\alpha\alpha,\mathrm{noise}}
     C_L^{TT,\mathrm{map}} W_L^{-2}.
\label{eqn:CalphaT_variance}
\end{equation}

Given an experimental configuration and a fiducial model with power spectra $C^{\alpha\alpha\text{,fiducial}}_L$ and $C^{\alpha T\text{,fiducial}}_L$, we can compute the expected signal-to-noise to which the birefringence signal could be detected by \cite{Caldwell:2011pu}:
\begin{eqnarray}
   (S/N)_{\alpha\alpha} &=& \left[ \sum_L
          \left(\frac{C^{\alpha\alpha\text{,fiducial}}}{\Delta
          \widehat C_L^{\alpha\alpha}}
     \right)^2\right]^{1/2}
\label{eqn:sigma_Aalphaalpha} \\
   (S/N)_{\alpha\mathrm{T}} &=& \left[ \sum_L \left(\frac{C^{\alpha
         T\text{,fiducial}}}{\Delta \widehat C_L^{\alpha T}}
     \right)^2\right]^{1/2}
\label{eqn:sigma_AalphaT}
\end{eqnarray}
In Table~\ref{tab:res} we report the minimal value of $G_{\gamma\phi}$ that could provide a detection such that $(S/N)=3$ for different CMB experimental configurations. We have found that current experimental limits provide a $95 \%$ C.L. bound of $G_{\gamma\phi}< 1 \cdot 10^{-15}~{\rm GeV}^{-1}$. Given this current upper limit, we can first note that a satellite experiment like LiteBIRD, ground-based experiments as Stage-III in the ``wide survey'' mode or the Simons Observatory in the LAT configuration, can improve current bounds on $G_{\gamma\phi}$ by more than one order of magnitude. However, both the Stage-III in ``deep survey'' mode, the Simons Observatory in SAT configuration and CMB-S4 can further improve these bounds: Stage-III deep and Simons Observatory SAT can improve by a factor $\sim 5$, while CMB-S4 can improve by one order of magnitude. The first lesson to learn, therefore, is that future experiments with better angular resolution will provide stringent constraints on the CB signal than experiments with more extensive sky coverage but a broader angular beam. Secondly, we can note that the most robust constraints will come from the $\alpha T$ cross-correlations instead of $\alpha \alpha$. This depends on the fact that while $C_L^{\alpha \alpha}$ scales as $G_{\gamma\phi}^2$, $C_L^{\alpha T}$ depends linearly from $G_{\gamma\phi}$. The $\alpha T$ correlations will, therefore, be extremely important in constraining small couplings. Finally, we see that the constraints on $G_{\gamma\phi}$ are weakly dependent on the choice of the potential with $n=2$ or $n=3$.

\begin{table}
\begin{center}
\scalebox{0.9}{%
\begin{tabular}{l|cc|cc}
\toprule
\horsp
Experiment \vertsp $G_{\gamma\phi}$ \vertsp $G_{\gamma\phi}$ \vertsp $G_{\gamma\phi}$  \vertsp $G_{\gamma\phi}$ \\
 \vertsp $\alpha\alpha$, $n=2$ \vertsp $\alpha T$, $n=2$ \vertsp  $\alpha\alpha$, $n=3$ \vertsp $\alpha T$, $n=3$\\
%\vertsp $n=2$ \vertsp $n=3$\\
% \vertsp $n=2$ model \vertsp $n=2$ model \vertsp $n=3$ model \vertsp $n=3$ model\\
\hline
\morehorsp
LiteBIRD      & $0.145$ & $0.078$& $0.137$ & $0.087$\\
\morehorsp
S3wide      & $0.11$ & $0.068$& $0.11$ & $0.077$\\
\morehorsp
S3deep      & $0.0345$ & $0.034$& $0.033$ & $0.038$\\
\morehorsp
Simons Obs. SAT      & $0.031$ & $0.027$& $0.03$ & $0.029$\\
\morehorsp
Simons Obs. LAT      & $0.085$ & $0.052$& $0.08$ & $0.058$\\
\morehorsp
CMB-S4      &$0.014$ & $0.0088$ & $0.0135$ & $0.0096$ \\
\bottomrule
\hline
\end{tabular}
}
\end{center}
\caption{Forecasts for the minimal value of the coupling $G_{\gamma\phi}$, in units of $10^{-15} {\rm GeV}^{-1}$, such that future experiments can detect CB with a signal-to-noise ratio of  $(S/N)=3$.}
\label{tab:res}
\end{table}

\begin{table}
\begin{center}
\scalebox{0.9}{
\begin{tabular}{l|cc}
\toprule
\horsp
Experiment \vertsp $G_{\gamma\phi}$ \vertsp $G_{\gamma\phi}$\\
% \vertsp |$n=3$ model - $n=2$ model|  \vertsp |$n=3$ model - $n=2$ model|\\
\vertsp $\alpha \alpha$ \vertsp $\alpha T$\\
\hline
\morehorsp
LiteBIRD      & $0.44$ & $0.042$\\
\morehorsp
S3wide      & $0.34$ & $0.038$\\
\morehorsp
S3deep      & $0.205$ & $0.019$\\
\morehorsp
Simons Obs. SAT      & $0.095$ & $0.015$\\
\morehorsp
Simons Obs. LAT      & $0.25$ & $0.029$\\
\morehorsp
CMB-S4      &$0.043$ & $0.005$\\
\bottomrule
\hline
\end{tabular}
}
\end{center}
\caption{Forecasts for the minimal value of the coupling $G_{\gamma\phi}$, in units of $10^{-15} {\rm GeV}^{-1}$, such that future experiments could discriminate between EDE models proposed to resolve the Hubble tension, with potentials with $n=2$ and $n=3$, with a signal-to-noise ratio of $(S/N)=3$.}
\label{tab:diff}
\end{table}

As discussed in the previous section, the CB angular spectra for the $n=2$ and $n=3$ EDE models are significantly different in shape, especially at large values of $L$ and for the $\alpha T$ channel. It is therefore interesting to investigate what is the minimal value of the coupling $G_{\gamma\phi}$ that could let future experiments to discriminate between these two models with a signal-to-noise ratio of $S/N\sim 3$.

We report the results of this analysis in Table~\ref{tab:diff} where we indeed show the values of the minimal coupling needed for the experiment for discriminating between a model with $n=2$ and a model with $n=3$ with $S/N=3$. As we can see, in this case, the use of the $\alpha T$ correlation will be even more critical. This can be clearly explained by the significantly different shapes for the $\alpha T$ theoretical spectra between the two models. 

%%%%%%%%%%%%%%%%%%%%%%%%%%%%%%%%%%%%%%%%%%%%%%%%%%%%%%%%%%%%%%%%%%%%%
{\it Conclusions.}---In this {\it Letter} we computed the expected angular power spectra of the rotation angle $\alpha$ and of the cross-correlation with temperature $\alpha T$ for the EDE models proposed in \cite{Poulin:2018cxd} as a solution to the current Hubble tension. We also included a new effect, due to fluctuations in the distance to last scattering, which impact the EDE models and may be expected to contribute to the CB signal produced in late-time dark energy scenarios \cite{Gluscevic:2009mm}. We found that future experiments as LiteBIRD can improve current constraints on a possible EM coupling $G_{\gamma\phi}$ by one order of magnitude with respect to current bounds. Experiments as CMB-S4 can improve these bounds by a further order of magnitude, showing the importance of polarization measurements at small angular scales. Interestingly, we also found that the two proposed EDE models predict significantly different angular spectra, especially in the cross-correlation $\alpha T$. Future polarization measurements can therefore not only confirm EDE as a solution to the Hubble tension but also help in discriminating between different scenarios.

%%%%%%%%%%%%%%%%%%%%%%%%%%%%%%%%%%%%%%%%%%%%%%%%%%%%%%%%%%%%%%%%%%%%%
\acknowledgments
AM thanks the University of Manchester and the Jodrell Bank Center for Astrophysics for hospitality. LMC and AM are supported by TASP, iniziativa specifica INFN.

%%%%%%%%%%%%%%%%%%%%%%%%%%%%%%%%%%%%%%%%%%%%%%%%%%%%%%%%%%%%%%%%%%%%%
\bibliography{Hubble_Tension_PNGB_Birefringence}

\begin{thebibliography}{37}%
\makeatletter
\providecommand \@ifxundefined [1]{%
 \@ifx{#1\undefined}
}%
\providecommand \@ifnum [1]{%
 \ifnum #1\expandafter \@firstoftwo
 \else \expandafter \@secondoftwo
 \fi
}%
\providecommand \@ifx [1]{%
 \ifx #1\expandafter \@firstoftwo
 \else \expandafter \@secondoftwo
 \fi
}%
\providecommand \natexlab [1]{#1}%
\providecommand \enquote  [1]{``#1''}%
\providecommand \bibnamefont  [1]{#1}%
\providecommand \bibfnamefont [1]{#1}%
\providecommand \citenamefont [1]{#1}%
\providecommand \href@noop [0]{\@secondoftwo}%
\providecommand \href [0]{\begingroup \@sanitize@url \@href}%
\providecommand \@href[1]{\@@startlink{#1}\@@href}%
\providecommand \@@href[1]{\endgroup#1\@@endlink}%
\providecommand \@sanitize@url [0]{\catcode `\\12\catcode `\$12\catcode
  `\&12\catcode `\#12\catcode `\^12\catcode `\_12\catcode `\%12\relax}%
\providecommand \@@startlink[1]{}%
\providecommand \@@endlink[0]{}%
\providecommand \url  [0]{\begingroup\@sanitize@url \@url }%
\providecommand \@url [1]{\endgroup\@href {#1}{\urlprefix }}%
\providecommand \urlprefix  [0]{URL }%
\providecommand \Eprint [0]{\href }%
\providecommand \doibase [0]{http://dx.doi.org/}%
\providecommand \selectlanguage [0]{\@gobble}%
\providecommand \bibinfo  [0]{\@secondoftwo}%
\providecommand \bibfield  [0]{\@secondoftwo}%
\providecommand \translation [1]{[#1]}%
\providecommand \BibitemOpen [0]{}%
\providecommand \bibitemStop [0]{}%
\providecommand \bibitemNoStop [0]{.\EOS\space}%
\providecommand \EOS [0]{\spacefactor3000\relax}%
\providecommand \BibitemShut  [1]{\csname bibitem#1\endcsname}%
\let\auto@bib@innerbib\@empty
%</preamble>
\bibitem [{\citenamefont {Aghanim}\ \emph {et~al.}(2018)\citenamefont {Aghanim}
  \emph {et~al.}}]{Aghanim:2018eyx}%
  \BibitemOpen
  \bibfield  {author} {\bibinfo {author} {\bibfnamefont {N.}~\bibnamefont
  {Aghanim}} \emph {et~al.} (\bibinfo {collaboration} {Planck}),\ }\href@noop
  {} {\  (\bibinfo {year} {2018})},\ \Eprint {http://arxiv.org/abs/1807.06209}
  {arXiv:1807.06209 [astro-ph.CO]} \BibitemShut {NoStop}%
%%CITATION = ARXIV:1807.06209;%%
\bibitem [{\citenamefont {Riess}\ \emph {et~al.}(2019)\citenamefont {Riess},
  \citenamefont {Casertano}, \citenamefont {Yuan}, \citenamefont {Macri},\ and\
  \citenamefont {Scolnic}}]{Riess:2019cxk}%
  \BibitemOpen
  \bibfield  {author} {\bibinfo {author} {\bibfnamefont {A.~G.}\ \bibnamefont
  {Riess}}, \bibinfo {author} {\bibfnamefont {S.}~\bibnamefont {Casertano}},
  \bibinfo {author} {\bibfnamefont {W.}~\bibnamefont {Yuan}}, \bibinfo {author}
  {\bibfnamefont {L.~M.}\ \bibnamefont {Macri}}, \ and\ \bibinfo {author}
  {\bibfnamefont {D.}~\bibnamefont {Scolnic}},\ }\href {\doibase
  10.3847/1538-4357/ab1422} {\bibfield  {journal} {\bibinfo  {journal}
  {Astrophys. J.}\ }\textbf {\bibinfo {volume} {876}},\ \bibinfo {pages} {85}
  (\bibinfo {year} {2019})},\ \Eprint {http://arxiv.org/abs/1903.07603}
  {arXiv:1903.07603 [astro-ph.CO]} \BibitemShut {NoStop}%
%%CITATION = ARXIV:1903.07603;%%
\bibitem [{\citenamefont {Freedman}(2017)}]{Freedman:2017yms}%
  \BibitemOpen
  \bibfield  {author} {\bibinfo {author} {\bibfnamefont {W.~L.}\ \bibnamefont
  {Freedman}},\ }\href {\doibase 10.1038/s41550-017-0121} {\bibfield  {journal}
  {\bibinfo  {journal} {Nat. Astron.}\ }\textbf {\bibinfo {volume} {1}},\
  \bibinfo {pages} {0121} (\bibinfo {year} {2017})},\ \Eprint
  {http://arxiv.org/abs/1706.02739} {arXiv:1706.02739 [astro-ph.CO]}
  \BibitemShut {NoStop}%
%%CITATION = ARXIV:1706.02739;%%
\bibitem [{\citenamefont {Zhao}\ \emph {et~al.}(2017)\citenamefont {Zhao} \emph
  {et~al.}}]{Zhao:2017cud}%
  \BibitemOpen
  \bibfield  {author} {\bibinfo {author} {\bibfnamefont {G.-B.}\ \bibnamefont
  {Zhao}} \emph {et~al.},\ }\href {\doibase 10.1038/s41550-017-0216-z}
  {\bibfield  {journal} {\bibinfo  {journal} {Nat. Astron.}\ }\textbf {\bibinfo
  {volume} {1}},\ \bibinfo {pages} {627} (\bibinfo {year} {2017})},\ \Eprint
  {http://arxiv.org/abs/1701.08165} {arXiv:1701.08165 [astro-ph.CO]}
  \BibitemShut {NoStop}%
%%CITATION = ARXIV:1701.08165;%%
\bibitem [{\citenamefont {Di~Valentino}\ \emph
  {et~al.}(2017{\natexlab{a}})\citenamefont {Di~Valentino}, \citenamefont
  {Melchiorri}, \citenamefont {Linder},\ and\ \citenamefont
  {Silk}}]{DiValentino:2017zyq}%
  \BibitemOpen
  \bibfield  {author} {\bibinfo {author} {\bibfnamefont {E.}~\bibnamefont
  {Di~Valentino}}, \bibinfo {author} {\bibfnamefont {A.}~\bibnamefont
  {Melchiorri}}, \bibinfo {author} {\bibfnamefont {E.~V.}\ \bibnamefont
  {Linder}}, \ and\ \bibinfo {author} {\bibfnamefont {J.}~\bibnamefont
  {Silk}},\ }\href {\doibase 10.1103/PhysRevD.96.023523} {\bibfield  {journal}
  {\bibinfo  {journal} {Phys. Rev.}\ }\textbf {\bibinfo {volume} {D96}},\
  \bibinfo {pages} {023523} (\bibinfo {year} {2017}{\natexlab{a}})},\ \Eprint
  {http://arxiv.org/abs/1704.00762} {arXiv:1704.00762 [astro-ph.CO]}
  \BibitemShut {NoStop}%
%%CITATION = ARXIV:1704.00762;%%
\bibitem [{\citenamefont {Poulin}\ \emph {et~al.}(2018)\citenamefont {Poulin},
  \citenamefont {Boddy}, \citenamefont {Bird},\ and\ \citenamefont
  {Kamionkowski}}]{Poulin:2018zxs}%
  \BibitemOpen
  \bibfield  {author} {\bibinfo {author} {\bibfnamefont {V.}~\bibnamefont
  {Poulin}}, \bibinfo {author} {\bibfnamefont {K.~K.}\ \bibnamefont {Boddy}},
  \bibinfo {author} {\bibfnamefont {S.}~\bibnamefont {Bird}}, \ and\ \bibinfo
  {author} {\bibfnamefont {M.}~\bibnamefont {Kamionkowski}},\ }\href {\doibase
  10.1103/PhysRevD.97.123504} {\bibfield  {journal} {\bibinfo  {journal} {Phys.
  Rev.}\ }\textbf {\bibinfo {volume} {D97}},\ \bibinfo {pages} {123504}
  (\bibinfo {year} {2018})},\ \Eprint {http://arxiv.org/abs/1803.02474}
  {arXiv:1803.02474 [astro-ph.CO]} \BibitemShut {NoStop}%
%%CITATION = ARXIV:1803.02474;%%
\bibitem [{\citenamefont {Di~Valentino}\ \emph {et~al.}(2018)\citenamefont
  {Di~Valentino}, \citenamefont {Bøehm}, \citenamefont {Hivon},\ and\
  \citenamefont {Bouchet}}]{DiValentino:2017oaw}%
  \BibitemOpen
  \bibfield  {author} {\bibinfo {author} {\bibfnamefont {E.}~\bibnamefont
  {Di~Valentino}}, \bibinfo {author} {\bibfnamefont {C.}~\bibnamefont
  {Bøehm}}, \bibinfo {author} {\bibfnamefont {E.}~\bibnamefont {Hivon}}, \
  and\ \bibinfo {author} {\bibfnamefont {F.~R.}\ \bibnamefont {Bouchet}},\
  }\href {\doibase 10.1103/PhysRevD.97.043513} {\bibfield  {journal} {\bibinfo
  {journal} {Phys. Rev.}\ }\textbf {\bibinfo {volume} {D97}},\ \bibinfo {pages}
  {043513} (\bibinfo {year} {2018})},\ \Eprint
  {http://arxiv.org/abs/1710.02559} {arXiv:1710.02559 [astro-ph.CO]}
  \BibitemShut {NoStop}%
%%CITATION = ARXIV:1710.02559;%%
\bibitem [{\citenamefont {Di~Valentino}\ \emph
  {et~al.}(2017{\natexlab{b}})\citenamefont {Di~Valentino}, \citenamefont
  {Melchiorri},\ and\ \citenamefont {Mena}}]{DiValentino:2017iww}%
  \BibitemOpen
  \bibfield  {author} {\bibinfo {author} {\bibfnamefont {E.}~\bibnamefont
  {Di~Valentino}}, \bibinfo {author} {\bibfnamefont {A.}~\bibnamefont
  {Melchiorri}}, \ and\ \bibinfo {author} {\bibfnamefont {O.}~\bibnamefont
  {Mena}},\ }\href {\doibase 10.1103/PhysRevD.96.043503} {\bibfield  {journal}
  {\bibinfo  {journal} {Phys. Rev.}\ }\textbf {\bibinfo {volume} {D96}},\
  \bibinfo {pages} {043503} (\bibinfo {year} {2017}{\natexlab{b}})},\ \Eprint
  {http://arxiv.org/abs/1704.08342} {arXiv:1704.08342 [astro-ph.CO]}
  \BibitemShut {NoStop}%
%%CITATION = ARXIV:1704.08342;%%
\bibitem [{\citenamefont {Vattis}\ \emph {et~al.}(2019)\citenamefont {Vattis},
  \citenamefont {Koushiappas},\ and\ \citenamefont {Loeb}}]{Vattis:2019efj}%
  \BibitemOpen
  \bibfield  {author} {\bibinfo {author} {\bibfnamefont {K.}~\bibnamefont
  {Vattis}}, \bibinfo {author} {\bibfnamefont {S.~M.}\ \bibnamefont
  {Koushiappas}}, \ and\ \bibinfo {author} {\bibfnamefont {A.}~\bibnamefont
  {Loeb}},\ }\href {\doibase 10.1103/PhysRevD.99.121302} {\bibfield  {journal}
  {\bibinfo  {journal} {Phys. Rev.}\ }\textbf {\bibinfo {volume} {D99}},\
  \bibinfo {pages} {121302} (\bibinfo {year} {2019})},\ \Eprint
  {http://arxiv.org/abs/1903.06220} {arXiv:1903.06220 [astro-ph.CO]}
  \BibitemShut {NoStop}%
%%CITATION = ARXIV:1903.06220;%%
\bibitem [{\citenamefont {Pandey}\ \emph {et~al.}(2019)\citenamefont {Pandey},
  \citenamefont {Karwal},\ and\ \citenamefont {Das}}]{Pandey:2019plg}%
  \BibitemOpen
  \bibfield  {author} {\bibinfo {author} {\bibfnamefont {K.~L.}\ \bibnamefont
  {Pandey}}, \bibinfo {author} {\bibfnamefont {T.}~\bibnamefont {Karwal}}, \
  and\ \bibinfo {author} {\bibfnamefont {S.}~\bibnamefont {Das}},\ }\href@noop
  {} {\  (\bibinfo {year} {2019})},\ \Eprint {http://arxiv.org/abs/1902.10636}
  {arXiv:1902.10636 [astro-ph.CO]} \BibitemShut {NoStop}%
%%CITATION = ARXIV:1902.10636;%%
\bibitem [{\citenamefont {Knox}\ and\ \citenamefont
  {Millea}(2019)}]{Knox:2019rjx}%
  \BibitemOpen
  \bibfield  {author} {\bibinfo {author} {\bibfnamefont {L.}~\bibnamefont
  {Knox}}\ and\ \bibinfo {author} {\bibfnamefont {M.}~\bibnamefont {Millea}},\
  }\href@noop {} {\  (\bibinfo {year} {2019})},\ \Eprint
  {http://arxiv.org/abs/1908.03663} {arXiv:1908.03663 [astro-ph.CO]}
  \BibitemShut {NoStop}%
%%CITATION = ARXIV:1908.03663;%%
\bibitem [{\citenamefont {Adhikari}\ and\ \citenamefont
  {Huterer}(2019)}]{Adhikari:2019fvb}%
  \BibitemOpen
  \bibfield  {author} {\bibinfo {author} {\bibfnamefont {S.}~\bibnamefont
  {Adhikari}}\ and\ \bibinfo {author} {\bibfnamefont {D.}~\bibnamefont
  {Huterer}},\ }\href@noop {} {\  (\bibinfo {year} {2019})},\ \Eprint
  {http://arxiv.org/abs/1905.02278} {arXiv:1905.02278 [astro-ph.CO]}
  \BibitemShut {NoStop}%
%%CITATION = ARXIV:1905.02278;%%
\bibitem [{\citenamefont {Poulin}\ \emph {et~al.}(2019)\citenamefont {Poulin},
  \citenamefont {Smith}, \citenamefont {Karwal},\ and\ \citenamefont
  {Kamionkowski}}]{Poulin:2018cxd}%
  \BibitemOpen
  \bibfield  {author} {\bibinfo {author} {\bibfnamefont {V.}~\bibnamefont
  {Poulin}}, \bibinfo {author} {\bibfnamefont {T.~L.}\ \bibnamefont {Smith}},
  \bibinfo {author} {\bibfnamefont {T.}~\bibnamefont {Karwal}}, \ and\ \bibinfo
  {author} {\bibfnamefont {M.}~\bibnamefont {Kamionkowski}},\ }\href {\doibase
  10.1103/PhysRevLett.122.221301} {\bibfield  {journal} {\bibinfo  {journal}
  {Phys. Rev. Lett.}\ }\textbf {\bibinfo {volume} {122}},\ \bibinfo {pages}
  {221301} (\bibinfo {year} {2019})},\ \Eprint
  {http://arxiv.org/abs/1811.04083} {arXiv:1811.04083 [astro-ph.CO]}
  \BibitemShut {NoStop}%
%%CITATION = ARXIV:1811.04083;%%
\bibitem [{\citenamefont {Agrawal}\ \emph {et~al.}(2019)\citenamefont
  {Agrawal}, \citenamefont {Cyr-Racine}, \citenamefont {Pinner},\ and\
  \citenamefont {Randall}}]{Agrawal:2019lmo}%
  \BibitemOpen
  \bibfield  {author} {\bibinfo {author} {\bibfnamefont {P.}~\bibnamefont
  {Agrawal}}, \bibinfo {author} {\bibfnamefont {F.-Y.}\ \bibnamefont
  {Cyr-Racine}}, \bibinfo {author} {\bibfnamefont {D.}~\bibnamefont {Pinner}},
  \ and\ \bibinfo {author} {\bibfnamefont {L.}~\bibnamefont {Randall}},\
  }\href@noop {} {\  (\bibinfo {year} {2019})},\ \Eprint
  {http://arxiv.org/abs/1904.01016} {arXiv:1904.01016 [astro-ph.CO]}
  \BibitemShut {NoStop}%
%%CITATION = ARXIV:1904.01016;%%
\bibitem [{\citenamefont {Smith}\ \emph {et~al.}(2019)\citenamefont {Smith},
  \citenamefont {Poulin},\ and\ \citenamefont {Amin}}]{Smith:2019ihp}%
  \BibitemOpen
  \bibfield  {author} {\bibinfo {author} {\bibfnamefont {T.~L.}\ \bibnamefont
  {Smith}}, \bibinfo {author} {\bibfnamefont {V.}~\bibnamefont {Poulin}}, \
  and\ \bibinfo {author} {\bibfnamefont {M.~A.}\ \bibnamefont {Amin}},\
  }\href@noop {} {\  (\bibinfo {year} {2019})},\ \Eprint
  {http://arxiv.org/abs/1908.06995} {arXiv:1908.06995 [astro-ph.CO]}
  \BibitemShut {NoStop}%
%%CITATION = ARXIV:1908.06995;%%
\bibitem [{\citenamefont {Carroll}\ \emph {et~al.}(1990)\citenamefont
  {Carroll}, \citenamefont {Field},\ and\ \citenamefont
  {Jackiw}}]{Carroll:1989vb}%
  \BibitemOpen
  \bibfield  {author} {\bibinfo {author} {\bibfnamefont {S.~M.}\ \bibnamefont
  {Carroll}}, \bibinfo {author} {\bibfnamefont {G.~B.}\ \bibnamefont {Field}},
  \ and\ \bibinfo {author} {\bibfnamefont {R.}~\bibnamefont {Jackiw}},\ }\href
  {\doibase 10.1103/PhysRevD.41.1231} {\bibfield  {journal} {\bibinfo
  {journal} {Phys. Rev.}\ }\textbf {\bibinfo {volume} {D41}},\ \bibinfo {pages}
  {1231} (\bibinfo {year} {1990})}\BibitemShut {NoStop}%
%%CITATION = PHRVA,D41,1231;%%
\bibitem [{\citenamefont {Lue}\ \emph {et~al.}(1999)\citenamefont {Lue},
  \citenamefont {Wang},\ and\ \citenamefont {Kamionkowski}}]{Lue:1998mq}%
  \BibitemOpen
  \bibfield  {author} {\bibinfo {author} {\bibfnamefont {A.}~\bibnamefont
  {Lue}}, \bibinfo {author} {\bibfnamefont {L.-M.}\ \bibnamefont {Wang}}, \
  and\ \bibinfo {author} {\bibfnamefont {M.}~\bibnamefont {Kamionkowski}},\
  }\href {\doibase 10.1103/PhysRevLett.83.1506} {\bibfield  {journal} {\bibinfo
   {journal} {Phys. Rev. Lett.}\ }\textbf {\bibinfo {volume} {83}},\ \bibinfo
  {pages} {1506} (\bibinfo {year} {1999})},\ \Eprint
  {http://arxiv.org/abs/astro-ph/9812088} {arXiv:astro-ph/9812088 [astro-ph]}
  \BibitemShut {NoStop}%
%%CITATION = ASTRO-PH/9812088;%%
\bibitem [{\citenamefont {Molinari}\ \emph {et~al.}(2016)\citenamefont
  {Molinari}, \citenamefont {Gruppuso},\ and\ \citenamefont
  {Natoli}}]{Molinari:2016xsy}%
  \BibitemOpen
  \bibfield  {author} {\bibinfo {author} {\bibfnamefont {D.}~\bibnamefont
  {Molinari}}, \bibinfo {author} {\bibfnamefont {A.}~\bibnamefont {Gruppuso}},
  \ and\ \bibinfo {author} {\bibfnamefont {P.}~\bibnamefont {Natoli}},\ }\href
  {\doibase 10.1016/j.dark.2016.09.006} {\bibfield  {journal} {\bibinfo
  {journal} {Phys. Dark Univ.}\ }\textbf {\bibinfo {volume} {14}},\ \bibinfo
  {pages} {65} (\bibinfo {year} {2016})},\ \Eprint
  {http://arxiv.org/abs/1605.01667} {arXiv:1605.01667 [astro-ph.CO]}
  \BibitemShut {NoStop}%
%%CITATION = ARXIV:1605.01667;%%
\bibitem [{\citenamefont {Contreras}\ \emph {et~al.}(2017)\citenamefont
  {Contreras}, \citenamefont {Boubel},\ and\ \citenamefont
  {Scott}}]{Contreras:2017sgi}%
  \BibitemOpen
  \bibfield  {author} {\bibinfo {author} {\bibfnamefont {D.}~\bibnamefont
  {Contreras}}, \bibinfo {author} {\bibfnamefont {P.}~\bibnamefont {Boubel}}, \
  and\ \bibinfo {author} {\bibfnamefont {D.}~\bibnamefont {Scott}},\ }\href
  {\doibase 10.1088/1475-7516/2017/12/046} {\bibfield  {journal} {\bibinfo
  {journal} {JCAP}\ }\textbf {\bibinfo {volume} {1712}},\ \bibinfo {pages}
  {046} (\bibinfo {year} {2017})},\ \Eprint {http://arxiv.org/abs/1705.06387}
  {arXiv:1705.06387 [astro-ph.CO]} \BibitemShut {NoStop}%
%%CITATION = ARXIV:1705.06387;%%
\bibitem [{\citenamefont {Pogosian}\ \emph {et~al.}(2019)\citenamefont
  {Pogosian}, \citenamefont {Shimon}, \citenamefont {Mewes},\ and\
  \citenamefont {Keating}}]{Pogosian:2019jbt}%
  \BibitemOpen
  \bibfield  {author} {\bibinfo {author} {\bibfnamefont {L.}~\bibnamefont
  {Pogosian}}, \bibinfo {author} {\bibfnamefont {M.}~\bibnamefont {Shimon}},
  \bibinfo {author} {\bibfnamefont {M.}~\bibnamefont {Mewes}}, \ and\ \bibinfo
  {author} {\bibfnamefont {B.}~\bibnamefont {Keating}},\ }\href {\doibase
  10.1103/PhysRevD.100.023507} {\bibfield  {journal} {\bibinfo  {journal}
  {Phys. Rev.}\ }\textbf {\bibinfo {volume} {D100}},\ \bibinfo {pages} {023507}
  (\bibinfo {year} {2019})},\ \Eprint {http://arxiv.org/abs/1904.07855}
  {arXiv:1904.07855 [astro-ph.CO]} \BibitemShut {NoStop}%
%%CITATION = ARXIV:1904.07855;%%
\bibitem [{\citenamefont {Kamionkowski}(2009)}]{Kamionkowski:2008fp}%
  \BibitemOpen
  \bibfield  {author} {\bibinfo {author} {\bibfnamefont {M.}~\bibnamefont
  {Kamionkowski}},\ }\href {\doibase 10.1103/PhysRevLett.102.111302} {\bibfield
   {journal} {\bibinfo  {journal} {Phys. Rev. Lett.}\ }\textbf {\bibinfo
  {volume} {102}},\ \bibinfo {pages} {111302} (\bibinfo {year} {2009})},\
  \Eprint {http://arxiv.org/abs/0810.1286} {arXiv:0810.1286 [astro-ph]}
  \BibitemShut {NoStop}%
%%CITATION = ARXIV:0810.1286;%%
\bibitem [{\citenamefont {Gluscevic}\ \emph {et~al.}(2009)\citenamefont
  {Gluscevic}, \citenamefont {Kamionkowski},\ and\ \citenamefont
  {Cooray}}]{Gluscevic:2009mm}%
  \BibitemOpen
  \bibfield  {author} {\bibinfo {author} {\bibfnamefont {V.}~\bibnamefont
  {Gluscevic}}, \bibinfo {author} {\bibfnamefont {M.}~\bibnamefont
  {Kamionkowski}}, \ and\ \bibinfo {author} {\bibfnamefont {A.}~\bibnamefont
  {Cooray}},\ }\href {\doibase 10.1103/PhysRevD.80.023510} {\bibfield
  {journal} {\bibinfo  {journal} {Phys. Rev.}\ }\textbf {\bibinfo {volume}
  {D80}},\ \bibinfo {pages} {023510} (\bibinfo {year} {2009})},\ \Eprint
  {http://arxiv.org/abs/0905.1687} {arXiv:0905.1687 [astro-ph.CO]} \BibitemShut
  {NoStop}%
%%CITATION = ARXIV:0905.1687;%%
\bibitem [{\citenamefont {Pospelov}\ \emph {et~al.}(2009)\citenamefont
  {Pospelov}, \citenamefont {Ritz}, \citenamefont {Skordis}, \citenamefont
  {Ritz},\ and\ \citenamefont {Skordis}}]{Pospelov:2008gg}%
  \BibitemOpen
  \bibfield  {author} {\bibinfo {author} {\bibfnamefont {M.}~\bibnamefont
  {Pospelov}}, \bibinfo {author} {\bibfnamefont {A.}~\bibnamefont {Ritz}},
  \bibinfo {author} {\bibfnamefont {C.}~\bibnamefont {Skordis}}, \bibinfo
  {author} {\bibfnamefont {A.}~\bibnamefont {Ritz}}, \ and\ \bibinfo {author}
  {\bibfnamefont {C.}~\bibnamefont {Skordis}},\ }\href {\doibase
  10.1103/PhysRevLett.103.051302} {\bibfield  {journal} {\bibinfo  {journal}
  {Phys. Rev. Lett.}\ }\textbf {\bibinfo {volume} {103}},\ \bibinfo {pages}
  {051302} (\bibinfo {year} {2009})},\ \Eprint {http://arxiv.org/abs/0808.0673}
  {arXiv:0808.0673 [astro-ph]} \BibitemShut {NoStop}%
%%CITATION = ARXIV:0808.0673;%%
\bibitem [{\citenamefont {Yadav}\ \emph {et~al.}(2009)\citenamefont {Yadav},
  \citenamefont {Biswas}, \citenamefont {Su},\ and\ \citenamefont
  {Zaldarriaga}}]{Yadav:2009eb}%
  \BibitemOpen
  \bibfield  {author} {\bibinfo {author} {\bibfnamefont {A.~P.~S.}\
  \bibnamefont {Yadav}}, \bibinfo {author} {\bibfnamefont {R.}~\bibnamefont
  {Biswas}}, \bibinfo {author} {\bibfnamefont {M.}~\bibnamefont {Su}}, \ and\
  \bibinfo {author} {\bibfnamefont {M.}~\bibnamefont {Zaldarriaga}},\ }\href
  {\doibase 10.1103/PhysRevD.79.123009} {\bibfield  {journal} {\bibinfo
  {journal} {Phys. Rev.}\ }\textbf {\bibinfo {volume} {D79}},\ \bibinfo {pages}
  {123009} (\bibinfo {year} {2009})},\ \Eprint {http://arxiv.org/abs/0902.4466}
  {arXiv:0902.4466 [astro-ph.CO]} \BibitemShut {NoStop}%
%%CITATION = ARXIV:0902.4466;%%
\bibitem [{\citenamefont {Caldwell}\ \emph {et~al.}(2011)\citenamefont
  {Caldwell}, \citenamefont {Gluscevic},\ and\ \citenamefont
  {Kamionkowski}}]{Caldwell:2011pu}%
  \BibitemOpen
  \bibfield  {author} {\bibinfo {author} {\bibfnamefont {R.~R.}\ \bibnamefont
  {Caldwell}}, \bibinfo {author} {\bibfnamefont {V.}~\bibnamefont {Gluscevic}},
  \ and\ \bibinfo {author} {\bibfnamefont {M.}~\bibnamefont {Kamionkowski}},\
  }\href {\doibase 10.1103/PhysRevD.84.043504} {\bibfield  {journal} {\bibinfo
  {journal} {Phys. Rev.}\ }\textbf {\bibinfo {volume} {D84}},\ \bibinfo {pages}
  {043504} (\bibinfo {year} {2011})},\ \Eprint {http://arxiv.org/abs/1104.1634}
  {arXiv:1104.1634 [astro-ph.CO]} \BibitemShut {NoStop}%
%%CITATION = ARXIV:1104.1634;%%
\bibitem [{\citenamefont {Aguirre}\ \emph {et~al.}(2019)\citenamefont {Aguirre}
  \emph {et~al.}}]{Ade:2018sbj}%
  \BibitemOpen
  \bibfield  {author} {\bibinfo {author} {\bibfnamefont {J.}~\bibnamefont
  {Aguirre}} \emph {et~al.} (\bibinfo {collaboration} {Simons Observatory}),\
  }\href {\doibase 10.1088/1475-7516/2019/02/056} {\bibfield  {journal}
  {\bibinfo  {journal} {JCAP}\ }\textbf {\bibinfo {volume} {1902}},\ \bibinfo
  {pages} {056} (\bibinfo {year} {2019})},\ \Eprint
  {http://arxiv.org/abs/1808.07445} {arXiv:1808.07445 [astro-ph.CO]}
  \BibitemShut {NoStop}%
%%CITATION = ARXIV:1808.07445;%%
\bibitem [{\citenamefont {Abazajian}\ \emph {et~al.}(2016)\citenamefont
  {Abazajian} \emph {et~al.}}]{Abazajian:2016yjj}%
  \BibitemOpen
  \bibfield  {author} {\bibinfo {author} {\bibfnamefont {K.~N.}\ \bibnamefont
  {Abazajian}} \emph {et~al.} (\bibinfo {collaboration} {CMB-S4}),\ }\href@noop
  {} {\  (\bibinfo {year} {2016})},\ \Eprint {http://arxiv.org/abs/1610.02743}
  {arXiv:1610.02743 [astro-ph.CO]} \BibitemShut {NoStop}%
%%CITATION = ARXIV:1610.02743;%%
\bibitem [{\citenamefont {Matsumura}\ \emph {et~al.}(2013)\citenamefont
  {Matsumura} \emph {et~al.}}]{Matsumura:2013aja}%
  \BibitemOpen
  \bibfield  {author} {\bibinfo {author} {\bibfnamefont {T.}~\bibnamefont
  {Matsumura}} \emph {et~al.},\ }\href {\doibase 10.1007/s10909-013-0996-1} {\
  (\bibinfo {year} {2013}),\ 10.1007/s10909-013-0996-1},\ \bibinfo {note} {[J.
  Low. Temp. Phys.176,733(2014)]},\ \Eprint {http://arxiv.org/abs/1311.2847}
  {arXiv:1311.2847 [astro-ph.IM]} \BibitemShut {NoStop}%
%%CITATION = ARXIV:1311.2847;%%
\bibitem [{\citenamefont {Suzuki}\ \emph {et~al.}(2018)\citenamefont {Suzuki}
  \emph {et~al.}}]{Suzuki:2018cuy}%
  \BibitemOpen
  \bibfield  {author} {\bibinfo {author} {\bibfnamefont {A.}~\bibnamefont
  {Suzuki}} \emph {et~al.},\ }\bibfield  {booktitle} {\emph {\bibinfo
  {booktitle} {{17th International Workshop on Low Temperature Detectors (LTD
  17) Kurume City, Japan, July 17-21, 2017}}},\ }\href {\doibase
  10.1007/s10909-018-1947-7} {\bibfield  {journal} {\bibinfo  {journal} {J.
  Low. Temp. Phys.}\ }\textbf {\bibinfo {volume} {193}},\ \bibinfo {pages}
  {1048} (\bibinfo {year} {2018})},\ \Eprint {http://arxiv.org/abs/1801.06987}
  {arXiv:1801.06987 [astro-ph.IM]} \BibitemShut {NoStop}%
%%CITATION = ARXIV:1801.06987;%%
\bibitem [{\citenamefont {Arvanitaki}\ \emph {et~al.}(2010)\citenamefont
  {Arvanitaki}, \citenamefont {Dimopoulos}, \citenamefont {Dubovsky},
  \citenamefont {Kaloper},\ and\ \citenamefont
  {March-Russell}}]{Arvanitaki:2009fg}%
  \BibitemOpen
  \bibfield  {author} {\bibinfo {author} {\bibfnamefont {A.}~\bibnamefont
  {Arvanitaki}}, \bibinfo {author} {\bibfnamefont {S.}~\bibnamefont
  {Dimopoulos}}, \bibinfo {author} {\bibfnamefont {S.}~\bibnamefont
  {Dubovsky}}, \bibinfo {author} {\bibfnamefont {N.}~\bibnamefont {Kaloper}}, \
  and\ \bibinfo {author} {\bibfnamefont {J.}~\bibnamefont {March-Russell}},\
  }\href {\doibase 10.1103/PhysRevD.81.123530} {\bibfield  {journal} {\bibinfo
  {journal} {Phys. Rev.}\ }\textbf {\bibinfo {volume} {D81}},\ \bibinfo {pages}
  {123530} (\bibinfo {year} {2010})},\ \Eprint {http://arxiv.org/abs/0905.4720}
  {arXiv:0905.4720 [hep-th]} \BibitemShut {NoStop}%
%%CITATION = ARXIV:0905.4720;%%
\bibitem [{\citenamefont {Kamionkowski}\ \emph {et~al.}(2014)\citenamefont
  {Kamionkowski}, \citenamefont {Pradler},\ and\ \citenamefont
  {Walker}}]{Kamionkowski:2014zda}%
  \BibitemOpen
  \bibfield  {author} {\bibinfo {author} {\bibfnamefont {M.}~\bibnamefont
  {Kamionkowski}}, \bibinfo {author} {\bibfnamefont {J.}~\bibnamefont
  {Pradler}}, \ and\ \bibinfo {author} {\bibfnamefont {D.~G.~E.}\ \bibnamefont
  {Walker}},\ }\href {\doibase 10.1103/PhysRevLett.113.251302} {\bibfield
  {journal} {\bibinfo  {journal} {Phys. Rev. Lett.}\ }\textbf {\bibinfo
  {volume} {113}},\ \bibinfo {pages} {251302} (\bibinfo {year} {2014})},\
  \Eprint {http://arxiv.org/abs/1409.0549} {arXiv:1409.0549 [hep-ph]}
  \BibitemShut {NoStop}%
%%CITATION = ARXIV:1409.0549;%%
\bibitem [{\citenamefont {Frieman}\ \emph {et~al.}(1995)\citenamefont
  {Frieman}, \citenamefont {Hill}, \citenamefont {Stebbins},\ and\
  \citenamefont {Waga}}]{Frieman:1995pm}%
  \BibitemOpen
  \bibfield  {author} {\bibinfo {author} {\bibfnamefont {J.~A.}\ \bibnamefont
  {Frieman}}, \bibinfo {author} {\bibfnamefont {C.~T.}\ \bibnamefont {Hill}},
  \bibinfo {author} {\bibfnamefont {A.}~\bibnamefont {Stebbins}}, \ and\
  \bibinfo {author} {\bibfnamefont {I.}~\bibnamefont {Waga}},\ }\href {\doibase
  10.1103/PhysRevLett.75.2077} {\bibfield  {journal} {\bibinfo  {journal}
  {Phys. Rev. Lett.}\ }\textbf {\bibinfo {volume} {75}},\ \bibinfo {pages}
  {2077} (\bibinfo {year} {1995})},\ \Eprint
  {http://arxiv.org/abs/astro-ph/9505060} {arXiv:astro-ph/9505060 [astro-ph]}
  \BibitemShut {NoStop}%
%%CITATION = ASTRO-PH/9505060;%%
\bibitem [{\citenamefont {Carroll}(1998)}]{Carroll:1998zi}%
  \BibitemOpen
  \bibfield  {author} {\bibinfo {author} {\bibfnamefont {S.~M.}\ \bibnamefont
  {Carroll}},\ }\href {\doibase 10.1103/PhysRevLett.81.3067} {\bibfield
  {journal} {\bibinfo  {journal} {Phys. Rev. Lett.}\ }\textbf {\bibinfo
  {volume} {81}},\ \bibinfo {pages} {3067} (\bibinfo {year} {1998})},\ \Eprint
  {http://arxiv.org/abs/astro-ph/9806099} {arXiv:astro-ph/9806099 [astro-ph]}
  \BibitemShut {NoStop}%
%%CITATION = ASTRO-PH/9806099;%%
\bibitem [{\citenamefont {Ma}\ and\ \citenamefont
  {Bertschinger}(1995)}]{Ma:1995ey}%
  \BibitemOpen
  \bibfield  {author} {\bibinfo {author} {\bibfnamefont {C.-P.}\ \bibnamefont
  {Ma}}\ and\ \bibinfo {author} {\bibfnamefont {E.}~\bibnamefont
  {Bertschinger}},\ }\href {\doibase 10.1086/176550} {\bibfield  {journal}
  {\bibinfo  {journal} {Astrophys. J.}\ }\textbf {\bibinfo {volume} {455}},\
  \bibinfo {pages} {7} (\bibinfo {year} {1995})},\ \Eprint
  {http://arxiv.org/abs/astro-ph/9506072} {arXiv:astro-ph/9506072 [astro-ph]}
  \BibitemShut {NoStop}%
%%CITATION = ASTRO-PH/9506072;%%
\bibitem [{\citenamefont {Liddle}\ and\ \citenamefont
  {Scherrer}(1999)}]{Liddle:1998xm}%
  \BibitemOpen
  \bibfield  {author} {\bibinfo {author} {\bibfnamefont {A.~R.}\ \bibnamefont
  {Liddle}}\ and\ \bibinfo {author} {\bibfnamefont {R.~J.}\ \bibnamefont
  {Scherrer}},\ }\href {\doibase 10.1103/PhysRevD.59.023509} {\bibfield
  {journal} {\bibinfo  {journal} {Phys. Rev.}\ }\textbf {\bibinfo {volume}
  {D59}},\ \bibinfo {pages} {023509} (\bibinfo {year} {1999})},\ \Eprint
  {http://arxiv.org/abs/astro-ph/9809272} {arXiv:astro-ph/9809272 [astro-ph]}
  \BibitemShut {NoStop}%
%%CITATION = ASTRO-PH/9809272;%%
\bibitem [{\citenamefont {Blas}\ \emph {et~al.}(2011)\citenamefont {Blas},
  \citenamefont {Lesgourgues},\ and\ \citenamefont {Tram}}]{Blas:2011rf}%
  \BibitemOpen
  \bibfield  {author} {\bibinfo {author} {\bibfnamefont {D.}~\bibnamefont
  {Blas}}, \bibinfo {author} {\bibfnamefont {J.}~\bibnamefont {Lesgourgues}}, \
  and\ \bibinfo {author} {\bibfnamefont {T.}~\bibnamefont {Tram}},\ }\href
  {\doibase 10.1088/1475-7516/2011/07/034} {\bibfield  {journal} {\bibinfo
  {journal} {JCAP}\ }\textbf {\bibinfo {volume} {1107}},\ \bibinfo {pages}
  {034} (\bibinfo {year} {2011})},\ \Eprint {http://arxiv.org/abs/1104.2933}
  {arXiv:1104.2933 [astro-ph.CO]} \BibitemShut {NoStop}%
%%CITATION = ARXIV:1104.2933;%%
\bibitem [{\citenamefont {Allison}\ \emph {et~al.}(2015)\citenamefont
  {Allison}, \citenamefont {Caucal}, \citenamefont {Calabrese}, \citenamefont
  {Dunkley},\ and\ \citenamefont {Louis}}]{Allison:2015qca}%
  \BibitemOpen
  \bibfield  {author} {\bibinfo {author} {\bibfnamefont {R.}~\bibnamefont
  {Allison}}, \bibinfo {author} {\bibfnamefont {P.}~\bibnamefont {Caucal}},
  \bibinfo {author} {\bibfnamefont {E.}~\bibnamefont {Calabrese}}, \bibinfo
  {author} {\bibfnamefont {J.}~\bibnamefont {Dunkley}}, \ and\ \bibinfo
  {author} {\bibfnamefont {T.}~\bibnamefont {Louis}},\ }\href {\doibase
  10.1103/PhysRevD.92.123535} {\bibfield  {journal} {\bibinfo  {journal} {Phys.
  Rev.}\ }\textbf {\bibinfo {volume} {D92}},\ \bibinfo {pages} {123535}
  (\bibinfo {year} {2015})},\ \Eprint {http://arxiv.org/abs/1509.07471}
  {arXiv:1509.07471 [astro-ph.CO]} \BibitemShut {NoStop}%
%%CITATION = ARXIV:1509.07471;%%
\end{thebibliography}%

%%%####################

\vfill
\end{document}